\documentclass[twocolumn,apj,iop,a4]{openjournal}  

\usepackage[breaklinks]{hyperref}

\usepackage{xcolor}

\begin{document}

\title{Optimization of Deep Learning Models for Radio Galaxy Classification}


\author{Philipp Denzel$^1$}
\author{Manuel Weiss$^1$}
\author{Elena Gavagnin$^1$}
\author{Frank-Peter Schilling$^{1*}$}

\affiliation{$^1$Zurich University of Applied Sciences ZHAW, 8400 Winterthur, Switzerland}

\email{frank-peter.schilling@zhaw.ch}



\begin{abstract}
Modern radio telescope surveys, capable of detecting billions of galaxies in wide-field surveys, have made manual morphological classification by humans impracticable. 
This applies in particular when the Square Kilometre Array Observatory (SKAO) becomes operable in 2027, which is expected to close an important gap in our understanding of the Epoch of Reionization (EoR) and other areas of astrophysics.
To this end, foreground objects, contaminants of the 21-cm signal, need to be identified and subtracted. Source finding and identification is thus an important albeit challenging task, especially in radio continuum images. In this work, we investigate the ability of AI and deep learning (DL) methods that have been previously trained on other data domains to localize and classify radio galaxies with minimal changes to their architectures.
 Various well-known pretrained neural network architectures for image classification and object detection are trained and fine-tuned and their performance is evaluated on a public radio galaxy dataset derived from the Radio Galaxy Zoo. A comparison between convolutional neural network (CNN)- and transformer-based algorithms is performed. The best performing architecture is systematically optimized, the influence of data preprocessing is studied, and an uncertainty estimation is performed by means of an ensemble analysis.
Radio source classification performance nearly comparable to the current leading customized models in the literature can be quite easily obtained using existing standard pretrained deep learning architectures, without modification and increase in complexity of the model architectures but rather adaptation of the data. Combining various transformations on replicated image channels seems particularly effective in this regard. Using an ensemble of models can also further improve performance to over 90\% accuracy, on par with top-performing models in the literature.
The results of this study can be transferred to the analysis of other survey data, e.g. from the Murchison Wide-field Array (MWA), and in the future be used to study the EoR by enabling galaxy foreground subtraction in continuum images from the SKAO.
\keywords{radio galaxy zoo, deep learning}
\end{abstract}



\maketitle

\section{Introduction}
\label{sec:intro}

The classification of galaxies is a fundamental task in astronomy, serving as a cornerstone for various astrophysical studies~\citep{Braun2015}. 
The morphology of galaxies provides valuable insights into the formation and evolution of galaxies over cosmic time~\citep{Conselice14}, the formation of large-scale structure and the cosmic distribution of dark matter~\citep{Reddick13}, or the accelerated expansion of the Universe~\citep{Pesce20}.
Upcoming surveys by, e.g., the Euclid mission\footnote{\url{https://www.esa.int/Science_Exploration/Space_Science/Euclid}} 
or the Square Kilometre Array Observatory (SKAO)\footnote{\url{https://www.skao.int/}}
will resolve billions of galaxies \citep{Scaramella22, Santos15}.
Their increased sensitivity yields better quality and farther-reaching observations, leading to more and higher resolution detections at redshifts up to $\sim5$~\citep{Tingay13, Trott20, Pritchard15, Koopmans15}.
In wide-field surveys, telescopes generally sacrifice angular resolution for aperture and distance, which complicates a precise classification due to the loss in image quality.
Radio low-frequency telescopes such as the SKA also mainly probe cold gas emissions, pulsars, quasars, and other radio-loud objects which are less noticeable in optical or near-optical observations~\citep{Shimabukuro2022}.
Thus, the shape of galaxies in these bandwidths is quite different from their typical Hubble-esque appearance in the optical, further increasing difficulty of the galaxy finding and classification tasks, e.g.~\citep{Hancock2018}.

Manual galaxy finding has long been abandoned in favour of algorithmic approaches based on classical computer-vision techniques such as thresholding~\citep{Hopkins02}, 
edge detection~\citep{Mouhcine05}, watershed segmentation~\citep{Berry15}, wavelet transform~\citep{Ellien21}, 
or clustering~\citep{Johnston14}. 
Correspondingly, more and more galaxy discoveries await analysis or validation.
Several strategies to approach this challenge have already been explored such as crowd-sourcing labour-intensive tasks in citizen science projects where non professional volunteers perform the initial data selection and preparation~\citep{Lintott08, Lintott10}. However, while citizen science projects harness human intuition and the ability to quickly identify and process interesting features, they will struggle to scale to the unprecedented volume of data expected from next-generation surveys.

Machine learning and especially deep learning have been recognized as promising solutions to manage these vast amounts of data.
Neural networks have the ability to overcome the curse of dimensionality and generalize to more abstract concepts, thereby imitating human-level attention while leveraging machine-level processing speeds.
The proliferation of DL models and, in particular, pretrained vision models (PVMs) for representation learning such as DINO~\citep{ZhangDino2022},
EfficientFormer~\citep{EfficientFormer} or YOLOv8~\citep{YoloV8} has facilitated the application of deep learning in natural sciences.
Still, PVMs are underutilized in many scientific disciplines due to the often considerable distribution shift between domain-specific data and the training data of these pretrained models.
The overabundance, diversity, and inherent complexity pose additional hurdles in the adoption of such models by scientists.
In radio astronomy, it has been shown that PVMs often fall short of traditional supervised models trained from scratch solely on domain data~\citep{Lastufka24}.
ResNets~\citep{ResNet} pretrained on ImageNet~\citep{Deng2009} seem to be an exception here, as a common backbone of more complex architectures.
So, domain-specific custom DL architectures, e.g.~\citet{Wu2019, Burke19}, have been successfully employed as proof-of-concept in radio galaxy detection and classification, demonstrating performance still comparable to the latest top-performing models in terms of precision and average accuracy.
On the other hand, PVMs typically offer more robust predictions by leveraging learned representations across many domains.
They are particularly advantageous in optimizing model performance with limited labelled data (in a weakly supervised regime), often comprising only a few hundred or thousand images, which is a common scenario in astronomical applications.

Thus, this work focuses on training and evaluation of 
widely-used DL models
established within the DL literature and which have demonstrated outstanding performance on standard computer vision (CV) benchmark datasets, applied in the context of radio galaxy classification and detection.
The focus is on identifying best practices for adapting these models to domain-specific datasets, and vice versa.
Notably, the choice of data preprocessing techniques can significantly impact the effectiveness of a pretrained model in adapting to a new data domain.
Finally, most DL models give little to no feedback on the confidence in their decision and, if at all, only implicitly take into account the inherent spread in the data distribution.
We employ a practical approach to these issues in the form of model ensembling.

In the following sections, we discuss related work on radio galaxy classification (section \ref{sec:relwork}) and present the dataset, DL architectures as well as the data processing methods used (section \ref{sec:methods}).
The results for radio galaxy detection and classification are discussed in section \ref{sec:results}, followed by a summary and outlook (section \ref{sec:summary}).


\section{Related Work}
\label{sec:relwork}

Machine learning has been integral to the field of radio astronomy, enabling significant advancements in data analysis and interpretation.
Methods like Decision Trees~\citep{Fayyad1993,Ball2006}, 
Support Vector Machines~\citep{Zhang2004,Sadeghi2021}, 
or Random Forests~\citep{Cheng2020} 
used to outperform simple neural networks for radio source classification~\citep{Alger2018,Becker2019}.
Convolutional neural networks (CNNs) were the first to reveal a clear advantage over these more traditional methods~\citep{Alhassan2018}. So, astronomers started adapting well-known architectures, such as LeNet~\citep{Mohan2022} or AlexNet~\citep{Tang2019}, modifying them for their specific data domain~\citep{Lukic2018}; for a detailed comparison, see~\citet{Becker2021}.

A common problem in radio astronomy is a heavy skew in the data distribution towards object classes exhibiting Gaussian-like brightness distributions barely above the noise level.
This is often caused by a lack of resolution and leads to less robust models.
\citet{Aniyan2017} and \citet{Kresnakova2021} mitigated this problem of class imbalance using data augmentation techniques, including rotation or mixing of synthetically generated data, e.g. using generative adversarial neural networks~\citep{Hosenie2018}.
More recent investigations go beyond classical supervised learning methods into the field of semi-supervised or unsupervised learning approaches to overcome the shortage of accurately labelled data~\citep{Polsterer2015, Galvin2020, Tang2022, Slijepcevic2022, Slijepcevic2023, Hossain2023}

The classification and detection of radio galaxies with more complex DL models has been attempted many times, although almost exclusively trained from scratch.
\citet{Wu2019} presented \textit{CLARAN}, a DL model designed to classify radio galaxy morphologies.
It is based on the Faster R-CNN method, an object detection network based on region proposals~\citep{Ren2015}, trained on (a subset of) the Radio Galaxy Zoo Data Release 1 catalogue (RGZ DR1; see section~\ref{sec:dataset} or \cite{Banfield2015, Wong24}).
The study demonstrates the potential to handle the large volume of data expected from next-generation radio surveys, achieving high accuracy.

\citet{Riggi2022} proposed another R-CNN variant, Mask R-CNN, named \textit{caesar-mrcnn} (Compact And Extended Source Automated Recognition) for the detection of radio continuum sources.
Their model was trained on data surveys from the Australian SKA Pathfinder (ASKAP) telescope.
It aims to reduce the false detection rate and improves the association of multiple disjoint islands into physical objects, as is the case for strong active galactic nucleus outflows.
The model achieved a classification precision and recall above 90\% but only a 65\% reliability.
\citet{Riggi2023, Mostert2022, Zhang2022, Wang2021} tried to improve on these efforts using the more performant one-stage object detection networks YOLO~\citep{Redmon2016,YoloV5}
 which are favoured for many applications in industrial sectors.

\citet{Sortino2023} provides a comparative study of these and similar object detection models such as Masked R-CNN~\citep{he2017mask}, Detectron2~\citep{wu2019detectron2}, DETR~\citep{Carion2020}, and EfficientDet~\citep{tan2020efficientdet}, finding promising results of slightly above 90\% accuracy for radio astronomical images.
Their benchmarks yield YOLO as the overall best performing model, even when compared against transformer-based models~\citep{Vaswani2023} such as DETR.

Finally, \citet{Lastufka24} identifies the potential of using pretrained vision foundation models such as DINOv2~\citep{DINOv2} or AM-RADIO~\citep{Ranzinger23} in optical and radio astronomy applications.
They demonstrate a good adaption of these foundation models for classification and detection in the optical spectrum, but observe a degradation in performance especially for the classification task of radio galaxies.
This study highlights the challenges of distribution shift and the need for careful consideration to model selection, fine-tuning, and data handling when adapting PVMs to the domain of radio astronomy. Moreover, \cite{Cecconello24} recently provided a benchmark of pretrained models via semi-supervised and self-supervised approaches and demonstrated that domain-specific pretraining can be even more effective in comparison to traditional pretraining.

With exception of the latter studies, pretrained models are still not even considered for application in radio astronomy on the presumption of incompatibility of models trained on natural images with the domain of radio astronomy.
To identify best practices for the adoption of pretrained models, we therefore aim to investigate how DL model performance changes for different architectures with various (low-effort) data preprocessing, data augmentation, and fine-tuning techniques.


\section{Methods}
\label{sec:methods}

\subsection{Dataset}
\label{sec:dataset}

For this work, the publicly available Radio Galaxy Zoo (RGZ) \citep{Banfield2015} dataset was employed.
RGZ is one of the largest radio galaxy classification datasets with around $100,000$ radio source images taken from the surveys {\em Faint Images of the Radio Sky at Twenty Centimeters} (FIRST) \citep{Becker1995} and {\em Australia Telescope Large Area Survey} (ATLAS) \citep{Franzen2015}. 

RGZ is organized as an online citizen science project that enlists volunteers to classify radio sources and their host galaxies.
It bases its methodology on the original Galaxy Zoo citizen science project \citep{Lintott08, Lintott10}.
After 5.5 years of operation, more than $12,000$ registered users contributed to the first data release (RGZ DR1; \cite{Wong24}) by labelling over 2.29 million queries, yielding $99,146$ source classifications for FIRST and $583$ for ATLAS.

However, to train a (supervised) object detection model, not only a dataset with labelled classes must be available but also annotated bounding boxes, which specify the coordinates of the recognised objects in an image.
Such annotations are available only for a small subset of RGZ DR1, produced by~\citet{Wu2019}, here referred to as RGZ OD.

The RGZ DR1 dataset is separated into classes in terms of number of peaks and number of components.
The number of components is defined as the number of discrete radio components that a source encompasses, identified at the $4\sigma$ flux-density threshold.
The number of peaks refers to the count of bright peaks in a radio source that are detected.
This results in classes that are named after the combination of these two characteristics and written as ${[\# \text{components}]\_[\# \text{peaks}]}$, e.g $1\_1$.
For instance, a double-lobed radio galaxy with small angular extent and no radio core may be identified as a source with one component-two peaks (1\_2) or a two component-two peaks (2\_2) if the two lobes appear disconnected in the radio image, which directly translates to the Fanaroff-Riley classification (FRI and FRII;~\cite{Fanaroff1974}).

\begin{figure*}[t]
\centering
\includegraphics[width=0.99\textwidth]{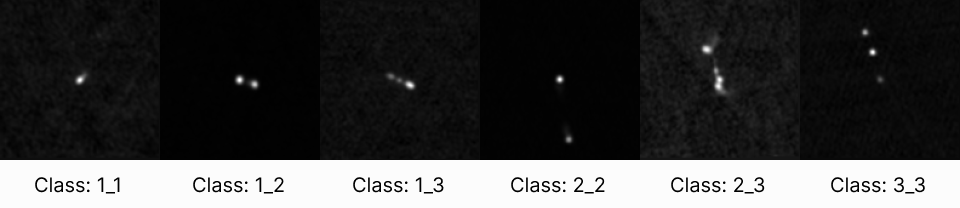}
\caption{Example images from the RGZ-OD dataset \citep{Wu2019}. The dataset is separated into classes in terms of number of peaks and number of components, where the former refers to the count of bright peaks that are detected, while the latter is defined as the number of discrete radio components. The resulting classes are named after the combination of the two as ${[\# \text{components}]\_[\# \text{peaks}]}$.
}
\label{fig.classes_rgz_od}
\end{figure*}

RGZ OD is the result of two filter rules that were applied to the RGZ DR1 dataset.
First, only samples that exceed a user-weighted consensus level (CL) of $\geq 0.6$ were selected, which should ensure that most radio sources are morphologically recognizable by humans.
Second, samples with more than four components and four peaks were disregarded to reduce class imbalance.
After these two rules, the RGZ OD contained in total $11,836$ samples in six classes: $1\_1,1\_2,1\_3,2\_2,2\_3$ and $3\_3$ (see figure~\ref{fig.classes_rgz_od}).

The RGZ OD image pixels represent aperture fluxes stored in a normalized range in units of $\mu$Jy/beam, from negative to positive values directly derived from the original cleaned and gridded images~\citep{Lukic2018}. Bounding boxes with respect to location and size were generated from the metadata considering the physical attributes defined in the RGZ DR1 dataset and converting sky coordinates into pixel coordinates.
In this work, the images provided in FITS
file format~\citep{Greisen2012} were used with the annotation information by~\citet{Wu2019}, to retain the accuracy of the intensity values before preprocessing.


\subsection{Deep Learning Architectures}
\label{sec:models}

The selected models for this study can roughly be grouped into two categories for each task (classification and object detection) respectively, convolutional and transformer-based networks. Pretrained weights are also available for the selected transformer-based networks, usually trained on ImageNet. From each category, we tested at least one well-established architecture to yield results independent from specific model features.

The following DL image classification architectures were considered: ResNet-50 \citep{ResNet} (here simply ResNet) is the 50-layer variant of the residual CNN architecture which won the ImageNet \citep{Deng2009} Large Scale Visual Recognition Challenge (ILSVRC) in 2015. Its layers optimize gradient flow through residual functions with respect to the layer inputs.
While ResNet is not considered state-of-the-art (SOTA) anymore, it is used as backbone in many other architectures. Thus, any performance enhancement on ResNet is likely to translate to other networks containing it as backbone.
EfficientNet \citep{EfficientNet} (EffNet) is a family of CNNs for computer vision published in 2019. Its key innovation is compound scaling, which uniformly scales all dimensions of depth, width, and resolution using a single parameter. Here, the variant EfficientNetV2-S \citep{EfficientNetV2} is used.
It yields superior performance to ResNet in many applications, and can be considered a SOTA among CNN architectures.
Vision Transformer~\citep{VisionTransformer} (ViT) and EfficientFormer \citep{EfficientFormer} (EffFor) are recent transformer models for computer vision. Analogous to their CNN-variants, they represent baseline and optimized variants of established transformer models.

For the object detection task, two architectures were used: YOLOv8-S \citep{YoloV8} (Yolov8) is a recent variant of the YOLO \citep{Redmon2016} family of single shot object detection models. It frames object detection as a regression task by spatially separating bounding boxes and associating probabilities to each detected image using a single neural network.
DINO~\citep{ZhangDino2022} (here: DINO; with a Swin, shifted window transformer, backbone~\citep{Liu2021}) is a transformer-based object detection model based on DETR~\citep{Carion2020} with various improvements.


%

\subsection{Preprocessing}
\label{sec:preproc}

Radio source images exhibit a significant amount of noise and show a narrow distribution of distinct intensity values. 
To ensure stable and fast training and increase performance, appropriate preprocessing of radio source images is essential \citep{Sharma2022}.

{\em Scaling} allows efficient and robust training and can prevent exploding or vanishing gradients. It can be used to constrain values within defined boundaries, e.g. between zero and one, or to distribute them more Gaussian-like. Here, two well-known scaling techniques are used: z-scale and min-max scale (normalization).
The z-scale algorithm was originally developed by the National Optical Astronomy Observatory and is implemented in the IRAF Framework \citep{Tody1986}. It is designed to display the intensity values near the median image, which is especially valuable for astronomical images, as they typically exhibit a peaked histogram in comparison to the background sky.
%
%

{\em Sigma clipping} is a common method used in astronomy to reduce noise during preprocessing~\citep{Lukic2018}, where values that deviate more than a certain number of standard deviations from the median or mean are discarded.

{\em Stretching} can be advantageous because, while the dynamic range in astronomical images may be large, the majority of values are near the noise level~\citep{White2007, Dumitrescu2022}. This also normalizes values across samples from different distributions and datasets, e.g., from different telescopes or standards on how the data is stored.
If this distinction is not a discriminatory factor, it can hinder the performance of a model. The following stretching methods are used: square root, power, power dist, linear, log, and sinh stretch.

{\em Dynamic Weight Conversion} takes the mean, standard deviation and skew of an image to dynamically extend a greyscale channel to three artificial colour channels \citep{Alrubaie2018}, in contrast to simply replicating it.

The preprocessing techniques were evaluated by means of a grid search in order to determine their influence on the performance of a basic YOLOv3 \citep{YoloV3} model without augmentation. All techniques were implemented using the Astropy library~\citep{Astropy2022}.
We observed a substantial increase in model performance from the z-scaling technique.
Further, the experiments show that sigma clipping as well as dynamic weights conversion can have a positive effect. On the other hand, the various stretching methods did not have a measurable impact on the results.
Thus, only z-scaling, sigma clipping, and dynamic weights conversion are reported for the experiments in section~\ref{sec:results}.

\subsection{Augmentation}
\label{sec:augmentation}

Data augmentation is a widely used strategy when data is scarce or imbalanced to mitigate overfitting or to increase the robustness of models. It includes applying label-preserving transformations to an input dataset in order to add more invariant examples \citep{Taylor2018}.
%
%
%
It alleviates the acute imbalance of the RGZ D1 dataset in the single-component, single-peak class compared to the other classes. Images are generally cropped to lie approximately in the centre of the image which creates a unfavourable weighting during training for offset source images as it is the case in the other classes.
%
%
The following augmentation techniques were evaluated for radio source detection and classification: Rotation (flipping in $90^\circ$ increments), translation, scaling, and shearing.



\section{Results}
\label{sec:results}

The models introduced in section \ref{sec:models} are trained and evaluated with a few basic variations of the training setup, including min-max (MM) scaling and z-scaling (ZS), the decision whether or not to use pretrained  weights (PT and NPT, respectively) on ImageNet \citep{Deng2009} for classification models and on COCO \citep{Lin2014} for object detection models.
Finally, a distinction is made between training on the imbalanced dataset or on a reduced but balanced dataset using undersampling (US).

Fine-tuning of the pretrained models (PT) was done by replacing the classification/projection heads, training the latter with frozen backbones first, and subsequently continue fine-tuning with unfrozen backbones after 5 epochs.
All models were trained for $300$ epochs (or until convergence) and a batch size of up to $32$.
The performance of the various DL model architectures is then evaluated for the radio source RGZ OD dataset~\citep{Wu2019}. Note that a limited-size dataset affects model performance of a CNN- in comparison to a transformer-based architecture differently but reflects the real-world use case in astronomy where labels are a rarity.

\subsection{Classification Task}
\label{sec:classification}

\begin{table}[t]
\centering
\begin{tabular}{@{}llll@{}}
\hline\hline
	{Model} & {Configuration} & {Acc. Top-1} & {Acc. Top-2}  \\ \hline
	ResNet         & ZS, NPT         & \textbf{83.38}          & \textbf{94.53}           \\
	ResNet         & ZS, NPT, US     & 81.42          & 93.90                     \\
	ResNet         & ZS, PT, US      & 81.42          & 94.47                     \\
	EffNet         & ZS, NPT         & 79.23          & 92.30                  \\
	EffNet         & ZS, PT          & 78.42          & 92.71                     \\
	EffNet         & ZS, PT, US      & 76.74          & 90.21                  \\
	EffFor         & ZS, NPT         & 76.60          & 89.46                 \\
	EffFor         & ZS, PT          & 75.48          & 89.56                 \\
	EffFor         & ZS, NPT, US     & 73.05          & 90.64                       \\
	ViT            & ZS, PT          & 59.47          & 77.91                  \\
	ViT            & ZS, NPT         & 58.46          & 77.41                    \\
	ViT            & MM, PT          & 52.99          & 70.72                     \\ \hline
\end{tabular}
\caption{Performance of classification models: Top-1 accuracy (Acc. Top-1) takes only the prediction with the highest score (output of the last hidden layer) into account, whereas Acc. Top-2 uses the two most likely predictions. ZS refers to the z-scaling preprocessing technique, MM to min-max scaling, US to undersampling to balance classes in the dataset, and (N)PT whether the model uses weights from pretraining (or not). All values are reported in per-centages.}
\label{tab:summary_performance_results_classification_benchmark}
\end{table}

Table \ref{tab:summary_performance_results_classification_benchmark} presents the three out of eight best working configurations of each examined classification model.
To provide comparability with previous works (see section~\ref{sec:relwork}), both top-1 and top-2 accuracies are provided. Moreover, since it is common to observe a confusion by most models (as well as human experts) between two pairs of classes (the class pairs $1\_2$, $1\_3$ and $2\_2$, $2\_3$, roughly corresponding to FR-1 and FR-2 galaxies, respectively), the top-2 accuracies show the general performance in absence of these confusions. 
%
To account for imbalanced classes in the dataset, the performance metrics are macro-weighted.

Surprisingly, ResNet-50 surpasses the performance of EfficientNet notably, with a maximum top-1 accuracy performance of~$83.38\%$.
On the other hand, unsurprisingly, the results confirmed our intuition that on the limited RGZ OD dataset transformer-based architectures cannot leverage their large number of parameters as an advantage.
Z-scaling showed to be more suitable for radio source images in comparison to min-max scaling for all applied models.
The experiments further indicate that from-scratch training slightly helps to optimise model performance, as opposed to fine-tuning pretrained models on ImageNet \citep{Deng2009}, which aligns with previous studies (see section~\ref{sec:intro}).
The model performance did generally not benefit from applying an undersampling filter to the imbalanced dataset.
Through inspection of misclassified samples, it can be observed that most confusion exists between class $1\_2$ and $1\_3$ as well as $2\_2$ and $2\_3$, which is also a distinction with which even human experts struggle most.








\subsection{Object Detection Task}
\label{sec:od_task}

Similar to the classification model benchmark experiments, the object detection models were trained on the RGZ OD dataset, with one difference in the preprocessing: Samples with multiple classes, which were excluded when training the classification models, are now considered. The performance results of the object detection benchmark experiments are listed in table~\ref{tab:summary_performance_results_object_detection_benchmark}.

\begin{table}[t]
\centering
\addtolength{\tabcolsep}{-2pt}
\begin{footnotesize}
\begin{tabular}{@{}lllllll@{}}
	\hline\hline
	{Model} & {Configuration} & {mAP} & {IOU} & {Acc.} &   ${mAP_{50}}$ & ${mAP_{75}}$ \\
	\hline
	DINO & MM, PT & \textbf{70.92} & 66.19 & \textbf{80.28} & \textbf{83.90} & \textbf{79.90} \\
	DINO & ZS, PT & 68.88 & 65.73 & 79.11 & 80.80 & 77.40 \\
	DINO & MM, PT, US & 67.80 & 64.83 & 76.60 & 80.90 & 77.00 \\
	DINO & ZS, PT, US & 66.02 & 64.76 & 74.47 & 78.30 & 74.90 \\
	DINO & MM, NPT & 64.71 & 67.32 & 78.70 & 79.20 & 74.70 \\
	DINO & ZS, NPT & 64.20 & 68.42 & 78.86 & 78.10 & 73.90 \\
	DINO & ZS, NPT, US & 63.05 & 70.88 & 73.01 & 76.80 & 72.70 \\
	DINO & MM, NPT, US & 61.18 & 67.52 & 72.21 & 75.10 & 70.10 \\
	YOLOv8 & ZS, PT, US & 55.75 & 88.08 & 71.68 & 69.80 & 66.00 \\
	YOLOv8 & ZS, PT & 52.72 & \textbf{89.08} & 70.09 & 66.10 & 62.00 \\
	YOLOv8 & MM, PT, US & 49.96 & 85.66 & 67.02 & 63.90 & 58.50 \\
	YOLOv8 & MM, PT & 47.41 & 85.97 & 69.09 & 62.40 & 55.50 \\
	YOLOv8 & ZS, NPT & 43.27 & 79.60 & 68.50 & 57.30 & 50.00 \\
	\hline
\end{tabular}
\end{footnotesize}
\caption{Performance of object detection models evaluated on the most common CV metrics (see subsection~\ref{sec:od_task}). All values are reported in per-centages.}
\label{tab:summary_performance_results_object_detection_benchmark}
\end{table}

The results show a different characteristic to the classification benchmark experiments.
Firstly, in terms of mean average precision (mAP) \citep{Padilla2020}, the YOLOv8 models could not outperform transformer-based models such as DINO. However, YOLOv8 achieved the best performance in terms of Intersection over Union (IoU).
This could be interpreted as DINO surpassing its CNN-based counterparts for the combination of object detection and source classification, while YOLOv8 performs best considering only the detection task.
On the other hand, as the bounding boxes in the RGZ OD dataset were generated automatically and always have a quadratic shape, the IoU has a systematic shift.
Lower IoU values could therefore also result from a deviation between inaccurate ground truth (GT) labelling and predicted boxes that are well-matched to the actual object.

In contrast to the classification experiments, using pretrained weights is beneficial. 
This could be explained by the comparatively higher complexity of object detection models and the ability to better generalize. In fact, the use of pretrained weights was the most influential factor leading to a higher mAP.
In addition, training the DINO models on images with min-max scaling resulted in higher performance compared to z-scaling.
Lastly, contrary to YOLOv8, DINO could not benefit from balancing the dataset with undersampling.

Inspecting samples misclassified by DINO models indicates a similar result as for the classification models with a strikingly higher confusion between class $1\_2$ and $1\_3$ as well as between $2\_2$ and $2\_3$.
Moreover, a frequent mistake involved misclassifying an actual source emission as background turning multi-component sources into $1\_1$.
This effect is most likely explained by the combination of an over-representation of the $1\_1$ class and general, inherent difficulty in dealing with noise in radio astronomy observations.

In summary, object detection models are generally more proficient at detecting radio sources, whereas the optimization of their classification performance is more challenging, leading to mixed results.

As the localization of radio sources can be comparatively accomplished with non-machine learning algorithmic finders, more emphasis was placed on further improving the classification task performance in the subsequent experiments.
%


\subsection{Model Uncertainty and Ensembling}
\label{sec:ensemble}


\begin{figure*}[t]
\centering
\includegraphics[width=0.99\textwidth]{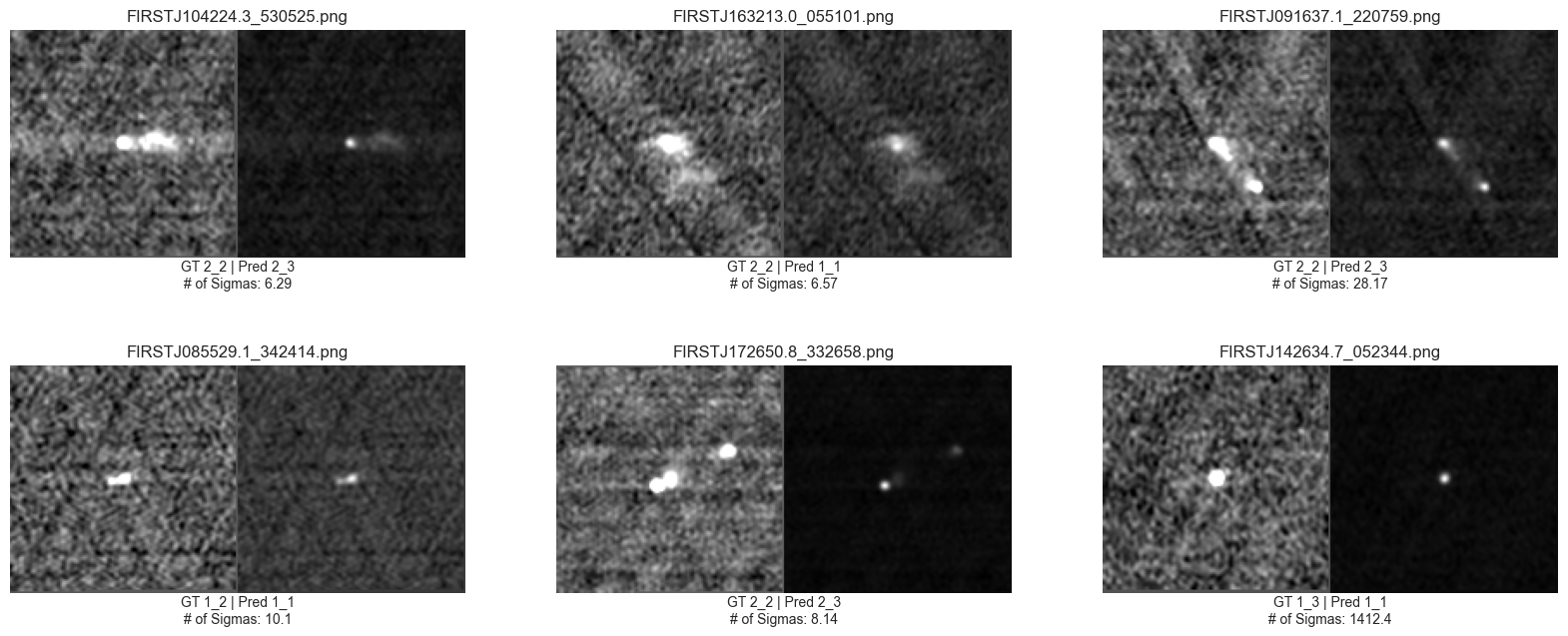}
\caption{High variance examples (for each panel, left: Z-scaled, right: unscaled).}
\label{fig.ensemble_resnet50_high_variance}
\end{figure*}


The ensemble method is applied to examine the (epistemic) uncertainty of the classification model predictions depending on the data quantity and quality at hand~\citep{Rahaman2021, Goran2020}.  
The softmax scores are used as the collaborative ensemble result from differently initialized and trained models, calculating their variance as statistical measure per class. 

The best performing classification model (ResNet) and settings from section~\ref{sec:classification} are used, training an ensemble of $30$ randomly initialized models on z-scaled radio sources, without pretrained weights.
The resulting uncertainty per class is in the range 0.041-0.062, with the exception of class $3\_3$, for which it is 0.022.
The average uncertainty is 0.0466.

Furthermore, the number of standard deviations of the difference between the mean value of the predictions and the ground truth was calculated for the test set.
$83.1\%$ of the samples are within one standard deviation. However, $4.6\%$ ($45$ samples) are above the six-sigma interval.
An excerpt of those high-variance samples is displayed in figure~\ref{fig.ensemble_resnet50_high_variance}.
Two phenomena can be observed: The first cluster of samples are actual misclassifications of the model. However, if the z-scaled representation on the left is compared with the unscaled representation on the right, it can be understood that some confusion of the model is attributable to different scaling methods. So, z-scaling seems to be advantageous for most of the samples but not for all.
The second cluster of misclassified images could have been mislabelled examples or at least difficult cases to which even humans cannot precisely assign a unique class.
These two clusters of errors can especially be observed for samples on which all 30 models agreed on the same different class from the ground truth.

%
Over all samples in the test set, on average, $27.66$ models out of $30$ agree on the predicted class with a standard deviation of 4. 
Further analysis shows that with an ensemble of three individually trained models, a possible top-1 accuracy of $90.68\%$ and a top-2 accuracy of $98.58\%$ can be achieved.

As discussed above, a major contributing cause to the model uncertainty is the inherent difficulty in distinguishing two pairs of classes (FR-I and FR-II galaxies, $1\_2$,$1\_3$ and $2\_2$,$2\_3$) which even domain experts have issues distinguishing at times. Moreover, image quality limitations induce an additional bias towards the trivial class ($1\_1$).




\subsection{Parameter Tuning}

The most promising model of the experiments in section \ref{sec:classification} (ResNet-50) is further fine-tuned with respect to specific choices for scaling, augmentation, optimizer, and preprocessing.





\begin{figure}[ht]
\centering
\includegraphics[width=1.0\columnwidth]{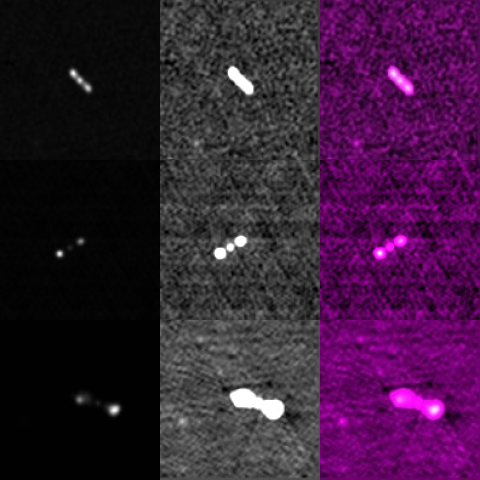}
\caption{Visual comparison of Min-Max scale (left), Z-scale (middle) and ZMZStack (right) for three example images.}
\label{fig.zmzstack}
\end{figure}

{\em Scaling:} The previous experiments (section \ref{sec:ensemble}) revealed that scaling has a significant impact on the performance of the classifier.
Examples were identified where the model classified noise as an additional component or peak due to the increase in brightness from z-scaling, or side lobes of weak intensity could not be distinguished from noise by the classifier without scaling. To remedy this shortcoming, two simple approaches were tested.
Firstly, tuning the contrast parameter of the z-scale implementation and secondly, the use of a z-scaled and a min-max scaled representation as different image channels.
The resulting preprocessed sample then contained three channels, two z-scaled and one min-max scaled in the centre, subsequently referred to as ZMZStack. A comparison with z-scaled and min-max scaled images is visualised in figure \ref{fig.zmzstack}. The highest performance increase was achieved when using ZMZStack, which increased the top-1 accuracy from $83.38\%$ to $86.12\%$ without further tuning.

{\em Augmentation:}  To account for possible distortions of the real distribution, rotation, scaling, translation and shearing were applied by a random factor $p$ between $0$ and $0.5$.
Each method was analysed individually and in combination. While scaling and translation were able to further increase performance, rotation had a neutral effect and shearing even a negative one. Ultimately, the highest improvement was achieved by a combination of scaling and translation with $p=0.25$, which further increased the top-1 accuracy to $88.35\%$.

{\em Optimizer: } Tuning the optimiser by strategy (SGD \citep{Ruder2017}, AdamW \citep{Loshchilov2019}, RMSProp \citep{Tieleman2012}), learning rate and learning rate scheduler could supplementary raise the performance level to $89.36\%$.

{\em Preprocessing: } Sigma median clipping could sightly improve the performance to $89.67\%$ top-1 accuracy.
However, in terms of top-2 accuracy, the performance declined to $97.47\%$. Sigma median clipping was subsequently only applied to the z-scaled channels of ZMZStack.

Finally, employing deeper models, e.g. ResNet-101 or ResNet-152, did not yield a significant increase in classification performance.
%
A summary of the incrementation steps is shown in table \ref{tab.summary_performance_results_classification_tuning}.

\begin{table}[t]
\centering
\begin{footnotesize}
\begin{tabular}{@{}llll@{}}
\hline\hline
{Model} & {Configuration} & {Acc. Top-1} & {Acc. Top-2} \\ \hline
ResNet         & Sigma Median Clipping         & \textbf{89.67}          &97.47                     \\
ResNet         & Optimiser         & 89.36          & \textbf{97.57}                \\
ResNet         & Augmentation      & 88.35          & 96.86                         \\
ResNet         & ZMZStack      & 86.12          & 94.43                               \\
ResNet         & Benchmark         & 83.38          &94.53                  \\ \hline
\end{tabular}
\end{footnotesize}
\caption{Performance of ResNet-50 after Parameter Tuning (reported analogous to figure~\ref{tab:summary_performance_results_classification_benchmark}).}
\label{tab.summary_performance_results_classification_tuning}
\end{table}

\section{Summary and Outlook}
\label{sec:summary}

Well-known DL architectures, which have demonstrated excellent performance in general computer vision tasks,
were applied to the task of detecting and classifying radio galaxies from the RGZ OD dataset, derived from the Radio Galaxy Zoo (RGZ D1) dataset.
The focus of this study was to investigate the data domain adaption of pretrained DL models, which up until recently has been believed to be inherently incompatible with the domain of radio astronomy data.
We found preprocessing the image data by means of scaling (z-scaling or min-max scaling) to be the most influential factors for model performance in classification.
Performance could be further improved by stacking image channels with two different scalings.
This suggests that further improvements in data adaption of pretrained models may be possible with suitable combinations of channel scalings.

While transformer-based PVMs such as DINO outperform other model architectures in terms of object detection, we obtained the best classification performance with lower-complexity, non-pretrained model architectures such as the ResNet-50.
This confirms the results by~\citet{Lastufka24} who observed similar behaviour. They have also demonstrated that this is not only due to the limited size of the dataset, as this behaviour is also observed with other, more diverse (but smaller) galaxy datasets in the radio spectrum. There, another limiting factor is the data quality, e.g. rather high noise levels due to instrumental limits imposed by current radio telescopes.

The performance of a ResNet-50 model could also be further increased in terms of average accuracy from $83.38\%$ to $89.67\%$ by means of scaling, data augmentation, tuning, and hyperparameter optimization.
Non-pretrained ResNet models exhibited marginally better performance, suggesting that learned representations on ImageNet have only favourable impact for models with a larger number of parameters.
In future studies, the domain adaption of models pretrained on smaller datasets should be investigated.
Using an ensemble of 30 models, we explored the model prediction's statistical uncertainty which was relatively low ($<1\sigma$) for the majority of the test set examples ($83\%$). The analysis of failure cases revealed actual misclassification, as well as potentially mislabelled training data.

We conclude that radio source classification performance nearly comparable to the current leading models in the literature (with an accuracy of above 90\%, \cite{Becker2021}) can be quite easily obtained using existing DL architectures, without modification and increase in complexity of the model architectures but rather adaptation of the data.
Indeed, while most models applied to the radio galaxy classification task in the literature are trained from scratch (no pre-trained weights) with elaborate, time-consuming modifications to the model architectures, our study suggests that models established in the CV community with large number of parameters including pre-trained weights can be used if data adaptation and fine-tuning is performed carefully.
Using an ensemble of models can also further improve performance to over 90\% top-1 accuracy on par with top-performing models in the literature with higher reliability.
With these data adaption methods, in combination with additional fine-tuning techniques such as cropping closer to the radio source, image rescaling to the original image size of the pre-training dataset, using larger projection heads, or more efficient pretraining strategies, we expect that pretrained models could yield superior performance compared to models trained from scratch with considerably less effort. A comparison to domain-specific pretrained models such as those by~\cite{Cecconello24} could lead to more insights in this regard and eventually lead to a complementary approach.

%
In the future, the results of this study can be transferred to surveys such as GLEAM-X (GaLactic and Extragalactic All-sky MWA Extended Survey~\citep{Gleam2015,Gleam2017}, based on observations with the Murchison Wide-field Array, or to surveys from the upcoming SKAO.

%
%


\bibliographystyle{aasjournal}
\bibliography{./references}

\end{document}